\documentstyle[12pt]{article}

\begin{document}
\title{Effect of Strong Magnetic Fields on the Equilibrium\\
 of a Degenerate Gas of Nucleons and Electrons}
\author{Choon-Lin Ho$^1$, V. R. Khalilov$^{1,\,2}$  and Chi Yang$^1$}
\date{\small $^1$Department of Physics, Tamkang University, Tamsui,
      Taiwan 251, R.O.C.\\
$^2$Department of Theoretical Physics, Physical Faculty,\\
Moscow State University, 119899, Moscow, Russia.\thanks{Permanent address.}}

\maketitle

\begin{abstract}

 We obtain the equations that define the equilibrium of a homogeneous
relativistic gas of neutrons, protons and electrons in a
constant magnetic field as applied to the conditions that probably occur
near the center of neutron stars. We compute the relative densities of
the particles at equilibrium and the  Fermi momentum of  electrons
in the strong magnetic field as function of the density of neutrons and
the magnetic field induction. Novel features are revealed as to the
ratio of the number of protons to the number of neutrons at equilibrium in
the presence of large magnetic fields.
\end{abstract}

\vskip 1 cm

\newpage

Significant interest in the behaviors of relativistic electrons in
a constant strong magnetic field was greatly stimulated by the
discovery of giant magnetic fields at the surface of neutron stars
\cite{Tr,Whe}.
Observations of hard X-rays from pulsar Hercules X-1 allowed one to
estimate the magnetic field induction at the pulsar surface to be of
the order of $ 10^{13}$ G. Such a magnetic field ``frozen'' in a
neutron star must become stronger and stronger reaching the ultrastrong
value of the order of
$10^{18} $ G in the central part of the neutron star \cite{TeDo}. Furthermore,
the fields of the order of $10^{23}$ G \cite{Vach} and, even possibly
$10^{33}$ G \cite{Dai} may exist at the electroweak
phase transition. It is worthy of note that the magnetic field in a neutron
star may be considered as a macroscopically uniform field with respect to
the characteristic scales, such as the Compton wavelengths, of constituent
particles of the neutron star. However, the field is extremely nonuniform
at the electroweak phase transition.  In any case, such a strong field
will affect in an essential way the behaviors of charged particles in the
star.

In this paper, we discuss the effects of strong constant magnetic fields on
the
chemical equilibrium of a degenerate gas of neutrons, protons, and electrons.
We assume in what follows that the gas is
spatially homogeneous due to the homogeneity of the magnetic field.

In a very strong magnetic field with induction $B$ higher than
$B_0=m^2 c^3/e\hbar=4.41\cdot 10^{13}$ G ($m$ and $e$ are the electron mass and
charge, respectively), electrons in the lowest Landau energy level
``move'' in the plane transverse to the magnetic field direction in
the region with characteristic dimensions of the order of
$\hbar(B_0/B)^{1/2}/mc $ (see, {\sl e.g.}
\cite{TeBag}).
We note that each electron energy level is degenerate with an infinite
multiplicity .  However, for the case under consideration the number of
quantum electron states is restricted by both the value of the
magnetic field induction and the size of the region in which the magnetic
field acts.

The first physical quantity which we must define is the chemical
potential $\mu_{\rm e}$, or the Fermi energy (if the gas temperature is equal
to zero), of relativistic electrons as a function of magnetic field induction
(see also \cite{Kha5}). The electron number density
$n_{\rm e}$ in the presence of a magnetic field ${\bf B}
=(0, 0, B)$ is defined by
\begin{eqnarray}
     n_{\rm e}={eB\over 4\pi^2\hbar^2 c}\sum_{n, \zeta}{\int_{-\infty}^{\infty}
     {dp\over \exp ((E_n-\mu_{\rm e})/\theta)+1}}\ ,
\label{eqn:1}
\end{eqnarray}
where
$ E_n=(p^2 c^2 + m^2 c^4 + 2eB\hbar cn)^{1/2}$
is the energy spectrum of a relativistic electron \cite{LandS};
$n=0, 1, 2, \ldots$
enumerates the Landau levels, $\zeta$ is the quantum number characterizing
the spin projection of an electron, $p$ is the electron momentum component
parallel to ${\bf B}$, and $\theta$ is the temperature. In what follows we
shall
express values of $\mu_{\rm e}$ in dimensionless units: $\mu=\mu_{\rm e}/mc^2$.
Since the chemical potential $\mu_0$ of an electron gas at $\theta=0$
and $B=0$ is related to $n_{\rm e}$ by
\begin{eqnarray}
     (\mu_0^2-1)^{3/2}=n_{\rm e}/n_0,\quad n_0=m^3 c^3/3\pi^2\hbar^3,
\label{eqn:4}
\end{eqnarray}
eq. (\ref{eqn:1}) may be considered as a relation between $\mu, \mu_0$
and $\theta$ at given $B$.
By  integrating with respect to $p$ in (\ref{eqn:1}), taking  allowance
for $E_{\rm max}=\mu_{\rm e}$ at $\theta=0$, it is easy to obtain:
\begin{eqnarray}
     {\sqrt 2n_{\rm e}\over 3n_0}\left({B_0\over B}\right)^{3/2}=
     \left((\mu^2-1){B_0\over 2B}\right)^{1/2}
     +2\sum_{n=1}^{n_{\rm max}}\left((\mu^2-1){B_0\over 2B}-n\right)^{1/2}.
\label{eqn:5}
\end{eqnarray}
The value $n_{\rm max}$ is determined by the condition
$ n_{\rm max}<(\mu^2-1)B_0/2B $.

It is seen from (\ref{eqn:5}) that,
if the following condition is satisfied:
\begin{eqnarray}
  (\sqrt 2n_{\rm e}/3n_0)\left(B_0/B\right)^{3/2}<1~,
\label{eqn:7}
\end{eqnarray}
then only the lowest (ground) level $n=0$ remains populated.
For a given $n_{\rm e}$, we can find from (\ref{eqn:5}) the relation
\begin{eqnarray}
     (\mu^2-1)^{1/2}=(2B_0/3B)(n_{\rm e}/n_0).
\label{eqn:9}
\end{eqnarray}

Let us now consider the conditions for chemical equilibrium of a
degenerate gas of protons
(p), neutrons (n) and electrons (e) in the presence  of large magnetic fields.
First we shall consider fields $B$ in the range
\begin{eqnarray}
     B_0^*\gg B\gg B_0,
\label{eqn:11}
\end{eqnarray}
where
$ B_0^*={m_{\rm p}^2c^3/e\hbar}=3.4 \cdot 10^6~B_0$, and
$m_{\rm p}$ is the proton mass. Magnetic field $B$ in this range is
``strong" for the electron but ``weak'' for the proton.  Hence  motion
of protons
in such a magnetic field can be considered as
quasiclassical
because the spacing between Landau levels for the proton under these
conditions is very small.
We assume further that the condition (\ref{eqn:7}) is also satisfied.
Then only the lowest energy level will be occupied by electrons. We also
suppose that the temperature $\theta$ is equal to zero, since for a typical
100-year old neutron star, its temperature is estimated to be $10^8$ K
(about $10$ keV), which can be considered cold as compared to the Fermi energy
(about $1000$ MeV) of the degenerate relativistic neutrons \cite{KW}.

We are interested in the reactions
in which the total density of baryons $n_{\rm b}=n_{\rm p}+n_{\rm n}$
is conserved and the electroneutrality condition of a gas $n_{\rm p}=n_{\rm e}$
is satisfied .
These processes are called the direct URCA processes \cite{KW,Wein,Latt,Bisn}.
Since  $n_{\rm b}$ is conserved, the total energy density
$\epsilon$ depends only on  $n_{\rm n}$:
\begin{eqnarray}
    \epsilon=\pi^{-2}\hbar^{-3}\!\!\int_0^{p_{\rm F}^{\rm n}}p^2
    (p^2c^2+m_{\rm n}^2 c^4)^{1/2}dp+\pi^{-2}\hbar^{-3}\!\!\int_0^{p_{\rm
F}^{\rm p}}p^2 (p^2c^2+m_{\rm p}^2c^4)^{1/2}dp
    +\epsilon_{\rm e}(p_{\rm F}^{\rm e}, B), 
\label{eqn:13}
\end{eqnarray}
where
\begin{eqnarray}
  p_{\rm F}^{\rm n}=(n_{\rm n}/n_{0\rm n})^{1/3}m_{\rm n}c,
  \quad
  p_{\rm F}^{\rm p}=(n_{\rm p}/n_{0\rm n})^{1/3}m_{\rm n}c,
  \nonumber             \\
  p_{\rm F}^{\rm e}=(2n_{\rm p}/3n_0)(B_0/B)mc,  \quad
   n_{0{\rm n}}=m_{\rm n}^3c^3/3\pi^2\hbar^3~,
\label{eqn:14}
\end{eqnarray}
and $\epsilon_{\rm e}$ is the energy density of the electrons.

At chemical equilibrium $d\epsilon/dn_{\rm n}=0$, or
$ \mu_{\rm n}=\mu_{\rm p}+\mu_{\rm e},$
where $\mu_{\rm n}, \mu_{\rm p}$  and $\mu_{\rm e}$  are the chemical
potentials of neutrons, protons and electrons respectively.
The chemical potential of a degenerate electron gas in the presence of
the magnetic field $B$ under the conditions assumed is
$\mu_{\rm e}$ given by formula (\ref{eqn:9}).  The quantity
$  p_{\rm F}^{\rm e}$ plays the role of
the Fermi momentum. More precisely, $p_{\rm F}^{\rm e}$
here is the ``Fermi momentum projection'' in the magnetic field direction
for electrons.

The chemical potentials of relativistic nucleons are given by
\begin{eqnarray}
    \mu_{\rm n}=m_{\rm n}c^2\left[1+(n_{\rm n}/n_{0{\rm
n}})^{2/3}\right]^{1/2},    \phantom{pppppp} \nonumber    \\
    \mu_{\rm p}=m_{\rm p}c^2\left[1+(n_{\rm p}/n_{0{\rm n}})^{2/3}
    (m_{\rm n}/m_{\rm p})^2\right]^{1/2},
\label{eqn:18}
\end{eqnarray}
for $B$ satisfying (\ref{eqn:11}).
The proton number density $n_{\rm p}$  as a function of $n_{\rm n}$  at
chemical equilibrium can be found from the following equation:
\begin{eqnarray}
    m_{\rm n}\left[1+(n_{\rm n}/n_{0{\rm n}})^{2/3}\right]^{1/2}=
    m_{\rm p}\left[1+(n_{\rm p}/n_{0{\rm n}})^{2/3}
    (m_{\rm n}/m_{\rm p})^2\right]^{1/2}
   \phantom{pppppp} \nonumber    \\
     +m\left[1+(2n_{\rm p}m_{\rm n}^3B_0)^2/
     (3n_{0{\rm n}}m^3B)^2\right]^{1/2}\!\!.\phantom{ppp}
\label{eqn:19}
\end{eqnarray}
Writing inequality (\ref{eqn:7}) in the form
\begin{eqnarray}
   n_{\rm p}/n_{0{\rm n}}<(3/\sqrt 2)\left(m/m_{\rm n}\right)^3
   \left(B/B_0\right)^{3/2}\approx
   3.4\cdot10^{-10}\left(B/B_0\right)^{3/2},
\label{eqn:20}
\end{eqnarray}
we see that, for $B_0^*\gg B\gg B_0$,
the inequality  $n_{\rm p}/n_{0{\rm n}}\ll 1$ must be satisfied.

We now turn to discuss equilibrium condition for the case of
ultrastrong magnetic fields
(with $B>B_0^*$) and very high densities
(with $n_{\rm p},\
n_{\rm n}> n_{0{\rm n}}$). The existence of such ultrastrong magnetic fields
near the center of neutron stars is allowed, or, at least, is not
forbidden in principle. In this case formula (\ref{eqn:19}) needs
to be modified because now the protons, just as the electrons, must occupy
their lowest Landau energy levels.  The limit of the proton density
under which all protons, and consequently all electrons, populate only
their respective ground levels is again given by (\ref{eqn:20}).  But now one could have $n_{\rm
p}/n_{0{\rm n}}> 1$. As in the case of electrons, the Fermi energy
for protons is given by
\begin{eqnarray}
    \mu_{\rm p}=m_{\rm p}c^2\left[1+(2n_{\rm p}B_0^*)^2/
    (3n_{0{\rm p}}B)^2\right]^{1/2}~.
\label{eqn:21}
\end{eqnarray}
Eq. (\ref{eqn:19}) is then replaced by
\begin{eqnarray}
    m_{\rm n}\!\left[1+(n_{\rm n}/n_{0{\rm n}})^{2/3}
    \right]^{1/2}\!\!=
    m_{\rm p}\!\left[1+(2n_{\rm p}B_0^*)^2/(3n_{0{\rm p}}B)^2
    \right]^{1/2}
   \phantom{ppppppppp} \nonumber    \\
    +m\!\left[1+(2n_{\rm p}m_{\rm n}^3B_0)^2/
    (3n_{0{\rm n}}m^3B)^2\right]^{1/2}\!\!\!.
\label{eqn:22}
\end{eqnarray}

Numerical solutions of eq. (\ref{eqn:19})  and (\ref{eqn:22})
subject to the constraint (\ref{eqn:20}) are given in
Fig.~1 ($B\le 10^5 B_0$ for eq. (\ref{eqn:19}), and $B\ge 10^7 B_0$ for
(\ref{eqn:22})).
These curves represent the normalized proton density number
$n_{\rm p}/n_{0{\rm n}}$ as a function of the normalized neutron
density number $n_{\rm n}/n_{0{\rm n}}$ at equilibrium
at various values of magnetic field induction.
The dotted curve gives the corresponding values in the absence of magnetic
fields.
The normalized Fermi momenta of electrons $p^{\rm e}_{\rm F}/mc$ as a
function of the
normalized neutron density number for these values of magnetic field are
given in Fig.~2.

One can see from Fig.~1 that results obtained here differ from analogous
results with zero magnetic field. Such a difference takes place at various
values of magnetic field induction  for some ranges both of low and high
particles densities.
Under the conditions assumed in this paper, for a given
value of ${n_{\rm n}/ n_{0{\rm n}}}$, the values
of ${n_{\rm p}/ n_{0{\rm n}}}$ in the presence of finite $B$'s are for
the most part higher than the corresponding value when the field is absent,
until the proton density reaches a value $(n_{\rm p}/ n_{0{\rm n}})_{\rm
cross}= (3B/ 2 B_0)^{3/2}~(m/m_{\rm
n})^3$, which is
the point of intersection between the curves with $B\ne 0$ and $B=0$.
The value of $(n_{\rm p}/ n_{0{\rm n}})_{\rm cross}$ is slightly less
than the upper limit given in (\ref{eqn:20}) for a given $B$.
Also, in the presence of ultrastrong magnetic field, there appear values of
densities for which
$n_{\rm p}>n_{\rm n}$.
These ranges of particle densities are of great interest since
in zero magnetic field the ratio ${n_{\rm p}/ n_{\rm n}}$ is always less than
unity, with a maximum equals 1/8 \cite{Wein}.
The functional form of (\ref{eqn:22}) implies that ${n_{\rm p}/ n_{0{\rm
n}}}$ is directly proportional to $B$.  This is evident in Fig.~1 from
the fact that curves corresponding to $B\ge 10^7 B_0$ are parallel to one
another, and are seperated by equal distance.

One may understand the behavior of these curves as follows.
Eq.~(\ref{eqn:14}) shows that higher magnetic field $B$ tends to lower the
Fermi momentum $p_{\rm F}^{\rm e}$ of the electrons, and hence the chemical
potential of electrons (chemical potential of protons as well, when $B>B_0^*$,
see (\ref{eqn:21})).  To maintain chemical equilibrium among the
particles at fixed value of $n_{\rm n}/n_{0{\rm n}}$, the chemical potentials
of electrons and protons have to be raised so that eq.(\ref{eqn:19}) or
(\ref{eqn:22}) is still satisfied.  This is achieved through the increase in
the density of
electrons, and hence the density of protons by neutrality condition assumed
here.  The shape of the curves, which consists mainly of segments of
straight lines in the $\log$-$\log$ plot, indicates power-law behavior between
the quantities $n_{\rm p}/n_{0{\rm n}}$ and $n_{\rm n}/n_{0{\rm n}}$. This can
also be easily understood.  Consider first the  situation in which $B>B_0^*$.
In the high density region where $n_{\rm n}/n_{0{\rm n}}\gg 1$, the
densities of electrons and protons are high enough that their Fermi
momenta satisfy
$p_{\rm F}^{\rm e}\gg mc$ and $p_{\rm F}^{\rm p}\gg m_{\rm p}c$.  Hence only
the second
term within each square-root in (\ref{eqn:22}) dominates.  After some algebra
we get
\begin{eqnarray}
{n_{\rm p}\over n_{0{\rm n}}}={3\over 4}~\left({m\over m_{\rm n}}\right)^2
~{B\over B_0}
~\left({n_{\rm n}\over n_{0{\rm n}}}\right)^{1/3}~.
\label{power1}
\end{eqnarray}
This shows that the straight lines on the upper right part of Fig.1 have slope
$1/3$.  For the region where $n_{\rm n}/n_{0{\rm n}}\ll 1$, the densities of
electrons and protons are lower such that $p_{\rm F}^{\rm p}\ll m_{\rm
p}c$, but still $p_{\rm F}^{\rm e}\gg mc$.
Eq.(\ref{eqn:22}) then reduces approximately to
\begin{eqnarray}
m_{\rm n}\left[1+{1\over 2}\left({n_{\rm n}\over n_{0{\rm
n}}}\right)^{2/3}\right]\approx m_{\rm p} + {2\over 3} {n_{\rm p}\over
n_{0{\rm n}}} {m_{\rm n}^3\over m^2} {B_0\over B}~.
\label{r2}
\end{eqnarray}
From this one has, after setting $m_{\rm n}\approx m_{\rm p}$,
\begin{eqnarray}
{n_{\rm p}\over n_{0{\rm n}}}={3\over 4}~\left({m\over m_{\rm n}}\right)^2
~{B\over B_0}
~\left({n_{\rm n}\over n_{0{\rm n}}}\right)^{2/3}~.
\label{power2}
\end{eqnarray}
The slopes of the corresponding line segments in Fig.1 are $2/3$.
We note here that (\ref{power1}) and (\ref{power2}) have the same coefficients
in front of the factor $n_{\rm n}/ n_{0{\rm n}}$, which mean the two line
segments in Fig.1 have the same intercept.
Similar consideration, when applied to (\ref{eqn:19}) for the case $B_0^*\gg
B\gg B_0$, leads also to the power law (\ref{power2}).  When $B=0$, we get the
power laws ${n_{\rm p}\over n_{0{\rm n}}}={1\over 8}{n_{\rm n}\over n_{0{\rm
n}}}$
and ${n_{\rm p}\over n_{0{\rm n}}}={1\over 8}({n_{\rm n}\over n_{0{\rm n}}})^2$
for high and low neutron density, respectively.

Fig.~2(a) shows that, for a given neutron density, the Fermi momentum of
electrons in the presence
of field $B<B_0^*$ decreases as $B$ increases.  It is not higher than the
corresponding value when $B=0$ until $(n_{\rm p}/ n_{0{\rm n}})_{\rm cross}$
is reached.  On the other
hand, one sees from Fig.~2(b) that electron Fermi momenta in the presence of
ultrastrong magnetic fields are independent of the magnetic field strength.
Furthermore, they equal the corresponding value in the absence of magnetic
field.  The former situation observed in Fig.~ 2(b) is just the
manifestation of the fact that, for a fixed $n_{\rm n} / n_{0\rm n}$ value,
$n_{\rm p}/ n_{0\rm n}$ is directly proportional to $B$ in this case, as
mentioned before, and the fact that $p^{\rm e}_{\rm F}$ depends only on their
ratio.  To understand the latter result, we simply observe that the equilibrium
equation (\ref{eqn:22}), expressed in terms of the Fermi momenta of the
nucleons,
and the electrons, is identical to the corresponding equation without magnetic
field.  This is because $p^{\rm n}_{\rm F}$ has the same form in both cases,
and while the corresponding expressions of $p^{\rm e}_{\rm F}$ and $p^{\rm p}_{\rm F}$ differ
in the two cases, the relation $p^{\rm e}_{\rm F} = p^{\rm p}_{\rm F}$ holds
well.  So for a fixed neutron density ({\it i.e.} fixed
$p^{\rm n}_{\rm F}$), both equations give the same solution for the value of
$p^{\rm e}_{\rm F}$, solved by different $n_{\rm p}/ n_{0\rm n}$ at different
$B$ when $B\ne 0$, of course.

The results obtained here may serve to give us hints as to
the chemical composition of the central region of neutron stars, if we suppose
that this region is composed of ``cold'' degenerate neutrons, protons and
electrons in the superstrong magnetic field. The assumption was also made
that all the electrons occupied only the lowest energy level when
$B\gg B_0$. If the gas temperature in this region is not equal
to zero, then we must introduce definite restrictions not only on the
quantity $n_{\rm e}$ but also on the gas temperature. There are several
aspects need to be considered.  Firstly, it is clear that at $\theta\ne 0$
the electrons can occupy the higher energy levels of transverse
(to the magnetic induction vector) motion of electrons due to thermal
excitations. Secondly, the electron-positron pair production due to
electron collisions in the ultradense hot medium can become noticeable.
Finally, the chemical potential of the electrons depends on the gas
temperature.

It is easy to see that the electrons will occupy the lowest energy level
of transverse motion at $B\gg B_0$ if 
\begin{eqnarray}
    {k_B\theta\over mc^2}\ll\left({2B\over B_0}\right)^{1/2}~,
\label{eqn:26}
\end{eqnarray}
where $k_B$ is the Boltzmann constant.
It follows from this inequality that, even at a very high temperature
of the order $m$ , electrons are still in the lowest
energy state. One must also estimate the effect of electron-positron pair
production by electron collisions in the presence of a magnetic field.
As a result of electron-positron pair production,
the equality $n_{\rm p}=n_{\rm e}$  will be violated but the
electroneutrality condition as a whole must be conserved. We estimate
that, even when  $B\gg B_0$, the
densities of electrons and positrons produced will be much less than the
initial density of electrons  $n_{\rm e}$ if the temperature satisfies the
strong inequality $k_B\theta/ mc^2\ll (n_{\rm e}B_0/3n_{0\rm n}B\ln 2)(m_{\rm
n}/ m)^3$. Thus the thermal effects of the electron-positron pair production
(even when $k_B\theta$ is of order $mc^2$) will not substantially affect
the chemical equilibrium of neutrons, protons and electrons over a
sufficiently wide interval of values $n_{\rm e}$.
We have
also made an estimate of the chemical potential of electrons which occupy the
lowest
Landau energy level as a function of temperature and found that, if inequality
(\ref{eqn:26}) is satisfied, then
the chemical potential in the form (\ref{eqn:9}) can still be employed.

Finally we remark that in this paper we consider only the case of ideal
gas model, as are usually done in the general analyses of basic properties of
compact stars \cite{KW,Wein}.
We see, however, that even such simple model
gives nontrivial and unexpected results.
So a careful analysis of more realistic models in
the presence of strong magnetic fields, and at finite temperatures, is
extremely needed to answer the question about the actual behavior
of such models.  Many factors need be considered.  For instance,
the density of matter near the center of a neutron
star is probably an inhomogeneous function of distance \cite{Latt,Bisn}, and
the problem of equilibrium in this case is significantly more complicated.
Also, one must consider the effect on the chemical equilibrium by the coulomb
interaction and the quantum exchange effect among the particles.
These consideration will be reserved for future investigations.

\vskip 2 truecm
\centerline{\bf Acknowlegdment}

This work is supported in part by the R.O.C. Grants number
NSC-85-2112-M-032-002 (C.-L.H.) and NSC-85-2112-M-032-005 (C.Y.).  V.R.K.
would like to thank the Department of Physics at the Tamkang University for
hospitality and financial support.

\vfil\eject

\vfil\eject
\centerline{\bf Figures Captions}
\vskip 20 pt

Fig.~1.\qquad  Plot of $\log (n_{\rm p} / n_{0\rm n})$ verses
$\log (n_{\rm n} / n_{0\rm n})$ (continuous curves) for values of $B/B_0$ at
(from
bottom to top): $10^2$,$~
10^3$,$~10^4$,$~10^5$,$~10^7$,$~10^9$,$~10^{11}$ and $~10^{13}$. The dotted
line corresponds to the curve with $B=0$.  Equation $\log (n_{\rm p} / n_{0\rm
n}) = \log (n_{\rm n} / n_{0\rm n})$ is indicated by the dashed line.

\vskip 8 pt

   Fig.~2 (a).\qquad Plot of the normalized Fermi momenta of electrons $p^{\rm
e}_{\rm F}/mc$ verses $n_{\rm n} / n_{0\rm n}$ for $B/B_0=10^4$
(upper curve) and $10^5$ (lower curve).  Dotted curve corresponds  to
$B=0$.
\vskip   8 pt

   Fig.~2 (b).\qquad Plot of the normalized Fermi momenta of electrons for
$p^{\rm  e}_{\rm F}/mc$ verses $n_{\rm n} / n_{0\rm n}$ for
$B/B_0=0$, $~10^7$, $~10^9$,$~10^{11}$ and $10^{13}$.

\end{document}